# Underlying magnetization responses of magnetic nanoparticles in assemblies

Mohammad Reza Zamani Kouhpanji[1, 2] Pieter B Visscher[4], and Bethanie J H Stadler[1, 3, *]

[1]Department of Electrical and Computer Engineering, [2]Department of Biomedical Engineering, [3]Department of Chemical Engineering and Materials Science, University of Minnesota, Minneapolis, USA. [4]Department of Physics and Astronomy, University of Alabama, USA.

* Corresponding author, Email: stadler@umn.edu, Tel.: +1(612) 626-1628.

**Abstract**

Magnetic nanoparticles (MNPs) have been proposed as an ultimate solution for diverse applications including nanomedicine and logic devices over decades. However, none has emerged revolutionary because realizing their magnetization response in an assembly is, surprisingly, still elusive. We employ our fast and universal magnetic characterization method, called the projection method, to underlie the reversible (irreversible) magnetization response of several assemblies. We then illustrate how the reversible (irreversible) magnetization response is correlated to the intrinsic properties (the coercivity and interaction fields) of the MNPs in an assembly. Our experimental observations indicate that the reversible magnetization does not solely depend on the interaction field, but it is indeed a function of the interaction field to coercivity ratio. Furthermore, for large porosities, the interaction field linearly changes with the porosity highlighting the predominant effects of dipole-dipole interactions on the interaction fields. However, at low porosities, the interaction field shows a nonlinear relationship with the porosity indicating the dipole fluctuations effects dominantly determine the interaction fields.

**Introduction**

Multifunctional magnetic nanoparticles (MNPs) are used in applications as diverse as therapeutics [1]–[4], logic devices [5], [6], magnetic sensors [7], [8], hard drives and random access memory [9], [10]. As a result, MNPs became fundamental nanomaterials for numerous research with different disciplines, including molecular biology and nanomedicine [11]–[15] as well as applied physics and nanostructured materials [16]–[22]. In all of these applications, it is critical to remotely control the magnetization response of the MNPs for ceasing and/or triggering sequential actions that require a comprehensive understanding of the magnetic characteristics (such as coercivity and interparticle interactions). For example, the MNPs have been proposed as building blocks for quantum storage [9], [18] and computing [23], [24], where the reversible (irreversible) magnetization response of the MNPs play a critical role on areal density, data stability, power consumption, and accuracy. Unfortunately, the high yielding nanofabrication processes of the MNPs do not allow the production of MNPs with identical magnetic characteristics leading to a distribution in the magnetic characteristics, and ultimately malfunctioning of the assembly. Furthermore, due to the inefficiency of the current magnetic characterization methods, a comprehensive demonstration of intrinsic magnetic properties' effects on the magnetic responses of the MNPs is still elusive. Consequently, the development of novel experiments for understanding the magnetic characteristics of the MNPs assemblies is in high demand.

First-order reversal curve (FORC) measurement is the most commonly used method for realizing the intrinsic magnetic properties of MNPs instead of the minor and major hysteresis loop measurements [25]. That is why the minor and major hysteresis loop measurements provide extremely limited information compared to the FORC measurements. Although the FORC measurement is considered as an exceptional method for the qualitatively and partially quantitatively illustrations of complex magnetic assemblies [26]–[30], it has three fundamental drawbacks. First, in taking two derivatives of the magnetization curves, noises are amplified in the FORC distribution. Smoothing can be used to reduce the noises, but it can also dramatically change or conceal the real features of the MNPs, as has been extensively studied previously [31]–[36]. Second, again taking two derivatives causes a loss of information from the magnetization data.

For example, some magnetization switching can occur immediately at the reversal field— for example, it is only a function of the reversal field— but some switching is completely independent of the initial magnetization states, for example, they are the only function of the applied field. Third, to reduce noise when measuring samples with weak magnetic moments, the field steps should be sufficiently small with large averaging times and large pause-times at the initial points (e.g.: at the saturation point and reversal field). These measurements can take several hours to days.

We proposed an ultrafast method for facile illustration of the intrinsic magnetic properties of MNPs that drastically suppresses the FORC measurement limitations [37]. This method relays on scanning the vicinity of the upper branch hysteresis loop (UBHL), as shown in Figure 1a. Intuitively, the projection method measures the upper branch of the hysteresis loop (UBHL) in the restricted green area as a function of two fields, the backward field ($H_b$) and the forward field ($H_f$). Theoretically, the derivative of the UBHL in the context of the projection method is

$$\frac{\partial M_{UBHL}(H_b)}{\partial H_b} = \left.\frac{\partial M(H_f, H_b)}{\partial H_f}\right)_{H_f=H_b} + \left.\frac{\partial M(H_f, H_b)}{\partial H_b}\right)_{H_f=H_b} \quad (1)$$
$$= RSF(H_b) + ISF(H_b)$$

where, the first term is the reversible magnetization (RSF) because it determines the spontaneous magnetization change with the applied field; and the second term is the irreversible magnetization (ISF) because it determines the residual magnetization between two sequential backward fields [25], [35], [37]. Figure 1 schematically illustrates the difference between them for the two different cases of the MNPs, non-interacting and interacting. For non-interacting MNPs with single coercivity, the spontaneous magnetization at the beginning of each curve is zero (the initial part is flat), see Figure 1b. Thus, the RSF is zero and ISF and UBHL are identical, behaving like a delta Dirac function centered at the coercivity. Similarly, if the coercivity has a Gaussian distribution, the ISF and UBHL are identical indicating a single Gaussian distribution. The situation for an interacting MNPs, on the other hand, is fundamentally different and can be categorized into two categories: I) the interaction field is weaker than the coercivity, and II) the interaction field is comparable or stronger than the coercivity. In the former, the interaction field is not sufficient to switch the MNPs, thus the RSF is fairly zero and ISF and UBHL are equivalent, see Figure 1c. In the latter, if the interaction field is sufficiently strong to switch the MNPs, the RSF will be non-zero leading to a deviation of the ISF from the UBHL. This fact is also valid for MNPs with randomly oriented magnetization, where the magnetization direction tends towards its equilibrium as field changes.

Moreover, the RSF and ISF can be used for determining the reversibility fraction as defined by the ratio of the reversible magnetization ($M_{rev}$) and irreversible magnetization ($M_{irr}$) as [35]

$$\text{Reversibility fraction} = \frac{M_{rev}}{M_{rev} + M_{irr}}(\%) = \frac{\int RSF(H)dH}{\int RSF(H_b)dH_b + \int ISF(H_b)dH_b} \quad (2)$$

Notice, if only the UBHL is measured (as it is done in the hysteresis loop measurements), the reversible and irreversible parts of the magnetization are unified leading to undetermined effects of the interaction field on the magnetization responses of the MNPs. In this case, since the derivative of the UBHL does not separate two independent physical properties of the MNPs, it is unable to draw a comprehensive picture of the magnetic characteristics of the MNPs, such as cases (b) and (c) in Figure 1. What is worse, even though the FORC measurement maps the whole area of hysteresis loops, its tedious data processing discards the RSF that causes an anomalous zero-field coercivity distribution [35].

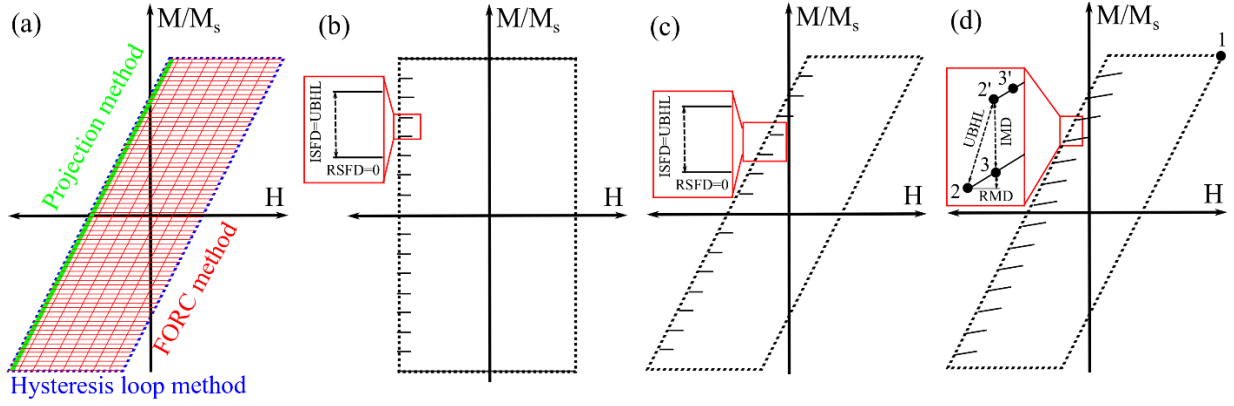

Figure 1: A schematic illustration of real characteristics for nanomagnets based on the projection method. (a) compares the collected data using the Projection method (green area), the FORC method (red area), and the hysteresis loop method (blue lines). The subfigure (a) shows the expected curves for non-interacting MNPs, (b) shows weakly interacting, and (c) shows strongly interacting.

According to our previous works [37], the projection method indeed provides two field distributions, given in Eq. (3) and (4), where they are equivalent to the integral of the FORC distribution on the reversal field ($H_r$, equivalent to the $H_b$) and the applied field (H, equivalent to the $H_f$) as follows

$$P_{H_f}(H_f) = \int_{-\infty}^{H_f} \rho(H_f, H_b) dH_b = -\frac{1}{2} \frac{\partial M(H_f, H_b)}{\partial H_f}\bigg|_{H_b=H_f} + \frac{1}{2}\frac{\partial M(H_f, H_b)}{\partial H_f}\bigg|_{H_b=-\infty}$$
$$= -\frac{1}{2}RSF + \frac{1}{2}\frac{\partial M_{lower}(H_f)}{\partial H_f} \quad (3)$$

$$P_{H_b}(H_b) = \int_{H_b}^{\infty} \rho(H_f, H_b) dH_f = -\frac{1}{2}\frac{\partial M(H_f, H_b)}{\partial H_b}\bigg|_{H_f=\infty} + \frac{1}{2}\frac{\partial M(H_f, H_b)}{\partial H_b}\bigg|_{H_f=H_b}$$
$$= 0 + \frac{1}{2}\frac{\partial M(H_f, H_b)}{\partial H_b}\bigg|_{H_f=H_b} = \frac{1}{2}ISF \quad (4)$$

In Eq. (3), $M_{lower}$ is the lower branch of the major hysteresis loop. The first term is the RSF because it represents the spontaneous variation of the magnetization with the field. Noted the first term is zero because for large $H_f$, M= +$M_s$, independently of $H_b$. The second term is the ISF because it determines the residual magnetization in the same field but started at different initial states. Since the projection method measures the UBHL, where the hysteresis loop is symmetry for MNPs. The $M_{lower}$ can thus be easily determined without any extra measurement, see Figure 1e.

According to Eq. (1), (3), and (4), the projection method provides the reversible (irreversible) magnetization response in addition to the intrinsic magnetic characteristics, the coercivity and interaction fields, all using a single measurement. Therefore, we use this powerful tool for underlining the reversible (irreversible) fraction in MNPs and how it is related to the intrinsic magnetic properties of the MNPs. Experimentally, we examine the projection method on several arrays of elongated MNPs, more specifically, magnetic nanowires (MNWs), in the direction of their easy axis to assure the reversible magnetization occurs only due to the interaction fields. After determining the intrinsic magnetic properties (the coercivity and interaction fields) in addition to the ISF and RSF, we elucidate the relation between the intrinsic magnetic properties of the MNPs and their reversible (irreversible) magnetization response.

**Experimental method**

Twelve magnetic nanowires (MNWs) arrays with different diameters, porosities, and materials were fabricated using the template-assisted electrodeposition technique, see ESI for more details. The nominal diameter (porosity) of the samples are as 20nm (12%), 30nm (0.5%), 50nm (1%), 80nm (15%), 100nm (2%), and 120nm (17%). In addition to those, six samples with 200nm diameter were prepared with three different porosities as 12%, 20%, and 60% using two different magnetic materials, nickel (Ni) and cobalt (Co). The magnetic anisotropy in the Ni MNW arrays is dominated by the shape anisotropy because Ni has a very small crystal anisotropy. Therefore, their easy axis is along with the Ni MNW arrays. However, the situation for the Co MNW arrays is different as they have a large crystal anisotropy. In this case, we electrodeposited the Co MNW arrays at pH > 6.5, as explained in the ESI, to assure that the crystal anisotropy and shape anisotropy are along the MNWs axis. The results for $P_{Hb}(H_b)$ and $P_{Hf}(H_f)$ distributions are calculated using Eq.(3) and Eq.(4) and provided in ESI.

**Results and discussions**

For small diameters or small porosities, both $P_{Hb}(H_b)$ and $P_{Hf}(H_f)$ distributions are fairly identical, meaning RSF=0, indeed they mirrored with the vertical axis, see ESI. These observations illustrate these MNWs arrays are non-interacting. The results for the $H_c$ and $H_{int}$ are given in Figure 2. The MNWs with smaller diameters have a larger coercivity and it decreases as the MNWs diameter increases. That is because the magnetic reversal mechanism in the smaller diameters occurs via the coherent rotation mechanism, in which all spins rotate spontaneously. However, as the diameter increases, the competition between the magneto-static energy and exchange energy leads to nucleation and propagation of a magnetic domain wall that requires less energy for switching the MNWs magnetization. As a result, the $H_c$ of the MNWs decreases as the diameter increases [38]–[40].

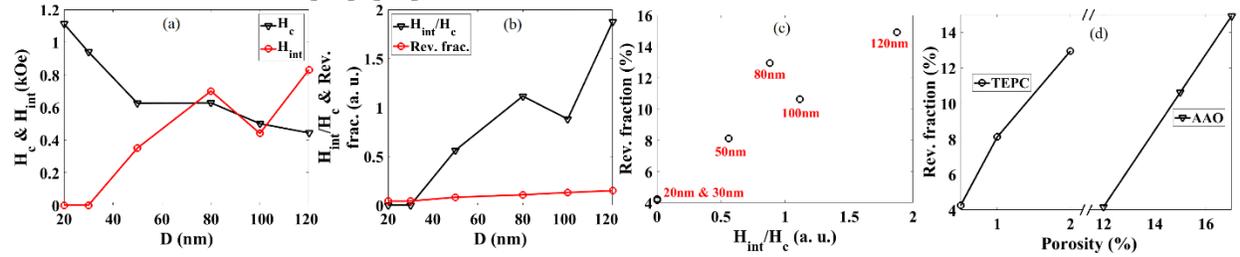

Figure 2: The magnetic characteristics for the Ni MNWs with different dimensions. (a) The coercivity and interaction fields, (b) and (c) are the interaction field ($H_{int}$) to coercivity ($H_c$) ratio and reversibility fraction.

As mentioned, the magnetic anisotropy of all the MNW arrays is along their axis, and the measurements were conducted in this direction. Thus, any reversibility in the magnetic response of the MNW arrays is due to the $H_{int}$. Practically, the $H_{int}$ does not cause reversibility unless they are comparable to the $H_c$ of the MNWs, see Figures 2b and 3d. The $H_{int}$ can be defined using the porosity while the $H_c$ is an intrinsic property of the MNWs, solely a function of their dimensions and composition. To examine this, our MNW array with the diameter of 20nm has a porosity of ~12% which is substantially larger than the ~0.5% and ~1% of the MNW arrays with diameters of 30nm and 50nm, respectively. Even though the 20nm and 30nm MNWs arrays have substantially different porosity, see Figure 2b, their reversible fraction is relatively similar and negligible. However, even though the 30nm and 50nm MNWs arrays have fairly similar porosity, their reversible fraction is different by a factor of 2X. This is due to the fact the $H_{int}$ in 20nm and 30nm MNW arrays are sufficiently smaller than their $H_c$ ($H_{int}/H_c$= ~0), see Figure 2c, while the $H_{int}$ in the 50nm MNWs arrays is comparable to their $H_c$ ($H_{int}/H_c$= 0.56).

Theoretically, the $H_{int}$ is a function of the MNW's volume, determining the amount of the magnetic compound, and the interwire distance ($D_{int}$) as follows [41]

$$H_{int} = \frac{\pi D^2 M_s}{8LD_{int}} \left(1 - \frac{D_{int}}{\sqrt{D_{int}^2 + L^2}}\right) \qquad (5)$$

where, D is the MNWs radius; L is the MNWs length; $D_{int}$ is the interwire distance between the MNWs; and $M_s$ is the magnetic moment of the MNWs. For the MNWs, since the length is much larger than the interwire distance, the term in the parenthesis of Eq. (5) approaches to one. As a result, the $H_{int}$ is proportional to $M_s D^2/LD_{int}$. This means that the reversibility fraction increases if: I) the radius of NWs increases while interwire distance and their type are similar, II) the interwire distance decreases while the NWs radius and type is similar, and III) the magnetic moment of the MNWs increases while their geometry and arrangement are identical.

To further elaborate on the effects of the porosity and magnetic composition, six types of the MNW arrays with a constant diameter (200nm) were prepared where the porosity and magnetic composition were tailored from ~12% to ~60%. Both Ni and Co MNWs arrays with the diameter of 200nm show a fairly constant $H_c$ regardless of their porosity. The minor variations in the $H_c$ are due to the nature of the templates, as the AAO templates have relatively more uniform pore size distribution compared to the TEPC templates. Even though the Co MNWs arrays have larger $H_{int}$ compared to the Ni MNWs arrays, Figure 3b, the Co MNWs arrays render the $H_{int}$ to $H_c$ ratio relatively similar to the Ni MNWs arrays, Figure 3c. That is because the Co MNWs have larger $H_c$ compared to the Ni MNWs arrays, Figure 3a, as a result of larger crystal anisotropy [39]. According to Figure 3b, both Ni and Co MNWs arrays demonstrate a linear relationship between the $H_{int}$ and the porosity, where the porosity is large. However, the $H_{int}$ for MNWs arrays with smaller porosity does not show a linear relationship. These observations highlight the $H_{int}$ is dominated by the dipole-dipole interaction among the MNWs with larger porosities [42], [43]. In contrast, as the porosity decreases, the $H_{int}$ does not follow a linear behavior meaning that the thermal fluctuations become dominant on the $H_{int}$, as previously observed [44], [45].

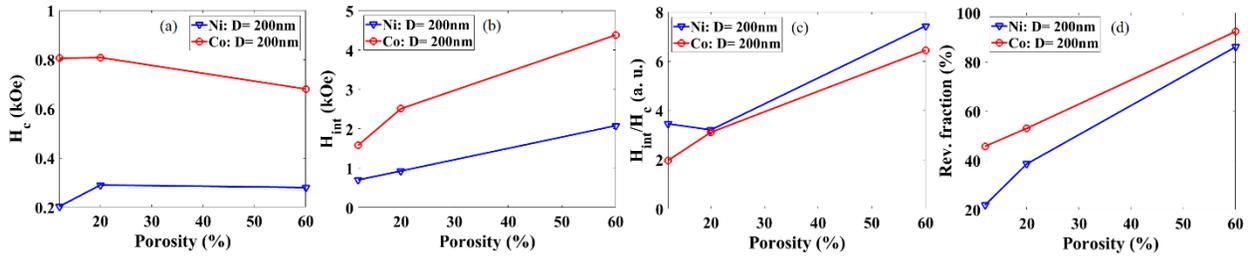

Figure 3: Illustrating the material type and porosity effects on (a) the coercivity, (b) the interaction field, (c) the interaction field to the coercivity ratio, and (d) the reversibility fraction.

The last, but not the least, interesting results of this analysis can be preserved from Figures 2d and 3d. They indicate that the reversible fraction linearly changes with the porosity regardless of the level of porosity or the MNWs composition. This finding highlights the porosity effects on the reversibility fraction that can shed light on the optimal design of the quantum storage and logic devices.

**Conclusion**

In this study, we introduced the capability of the projection method to elucidate the correlation between the density of the MNWs in an assembly and their reversible (irreversible) magnetization response, which is being lost in the hysteresis loop measurements and the FORC measurement. We showed that the interaction field ($H_{int}$) nonlinearly increases with the porosity if the porosity is small. This highlights that the $H_{int}$ of the MNWs arrays with smaller porosities dominated by the dipole fluctuations. However, as the porosity increases to large values, the $H_{int}$ represents a linear relationship with the porosity indicating the dipole-dipole coupling determines the $H_{int}$. This observation unambiguously resolved the discrepancy among the earlier studies as both linear and nonlinear behaviors were observed by different groups but it was not fully described due to the limitation of their methods. These findings highlight the robustness and reliability of the projection method for demonstrating the magnetic response of any magnetic nanoparticles that can boost the field of magnetism.